\documentclass[a4paper]{spie}  

\usepackage{amsmath,amsfonts,amssymb}
\usepackage{graphicx}
\usepackage[colorlinks=true, allcolors=blue]{hyperref}
\usepackage{comment}
\usepackage{float}
\usepackage{placeins}
\usepackage{subcaption}
\newcommand\dscript[2]{\bgroup#1\egroup_\bgroup#2\egroup}
\usepackage{caption}
\usepackage[acronym]{glossaries}
\glsdisablehyper

\title{Validation and extension of the PAWS Zemax model as a first step in the development of a Compact Arrayed Waveguide Stacked Multi-Object Spectrograph (CAWSMOS)}

\author[*a]{Fabienne Mangelsdorff}
\author[a]{Andreas Stoll}
\author[b]{Lars Zimmermann}
\author[b]{Stefan Lischke}
\author[b]{Riya M. Pattery}
\author[a,c]{Kalaga Madhav}
\author[a,c]{Martin M. Roth}
\affil[a]{Leibniz Institut für Astrophysik, Potsdam, Germany}
\affil[b]{Leibniz Institute for High Performance Microelectronics, Frankfurt (Oder), Germany}
\affil[c]{University of Potsdam, Potsdam, Germany}

\authorinfo{Send correspondence to Fabienne Mangelsdorff, fmangelsdorff@aip.de}
\makeglossaries
\begin{document}

\newacronym{paws}{PAWS}{Potsdam Arrayed Waveguide Spectrograph}
\newacronym{fpz}{FPZ}{Free Propagation Zone}
\newacronym{AWG}{AWG}{Arrayed Waveguide Grating}
\newacronym{cawsmos}{CAWSMOS}{Compact Arrayed Waveguide Stacked Multi-Object Spectrograph}
\newacronym{PSF}{PSF}{Point Spread Function}
\newacronym{SMF}{SMF}{Single Mode Fiber}
\newacronym{fwhm}{FWHM}{Full Width Half Maximum}
\maketitle

\begin{abstract}
The linear size of a bulk optical astronomical spectrograph scales with the diameter of the primary mirror of the corresponding telescope. As modern telescopes continue to increase in aperture size, miniaturization of the spectrograph becomes crucial beyond the offered advantages in terms of multifunctional integration and photon efficiency.
The presented concept of development for a \gls{cawsmos} includes the design of the \gls{AWG} chips, the stacking frame and cross-dispersion optics for the imaging of multiple spectral orders per \gls{AWG}.
It aims to reduce the cost and size of astronomical spectrographs while also improving efficiency. This will bring the \gls{AWG} technology closer to the realization of its full potential for ground based, airborne and spaceborne astronomical applications. 
Part of this development is an extension of the \gls{paws}. To do that, the Zemax model is compared to \gls{paws} calibration data, finding that the position can be matched with the measurement within 28 px in x and 18 px in y-direction. It also shows a discrepancy between the measured PSF size and the
Zemax model of around factor five, and that the detector area can only
partially accommodate a second chip.
\end{abstract}

\keywords{\gls{AWG}, Integrated Photonic Spectrograph, astrophotonics, arrayed waveguide grating, multi-object spectroscopy, Zemax modelling, point spread function, cross-dispersion, PAWS, H-band}

\section{INTRODUCTION}

\gls{paws} is a spectrograph for the astronomical H-band \cite{Hernandez2023} with a target resolving power of 15,000-30,000. From the telescope's focal plane light is fed via a photonic lantern into a \gls{SMF} which is butt-coupled to the \gls{AWG} \cite{Cvetojevic2009, Stoll2020}. The \gls{AWG} covers roughly the atmospheric H-band window (1450
nm - 1700 nm) and offers very low propagation losses and good fiber
compatibility by using a Silica-on-Silicon (SoS) material platform
with a refractive index contrast of 2\% \cite{Stoll2020}. In the spectrograph the \gls{AWG} serves as the primary dispersive element. Downstream of the \gls{AWG} a cryostat contains the bulk optics along with the cross-dispersing Echelle grating \cite{Hernandez2020, Hernandez2023}. To further exploit the potential of this technology, a new spectrograph called \gls{cawsmos} is being developed. \gls{cawsmos} shall extend \gls{paws} capabilities by including more input channels. As a first proof of concept, \gls{paws} is first extended by a single fiber then by a second \gls{AWG}. The goal is to stack multiple \gls{AWG}s and using multiple inputs per \gls{AWG} to investigate multiple objects at the same time with a compact spectrograph. Alternatively, light from a single multimode fiber, picking up the light from the telescope's focal plane, could be distributed to multiple different \gls{AWG} via a photonic lantern providing different ranges of wavelengths. 

\begin{figure}
    \centering
    \includegraphics[width=0.8\linewidth]{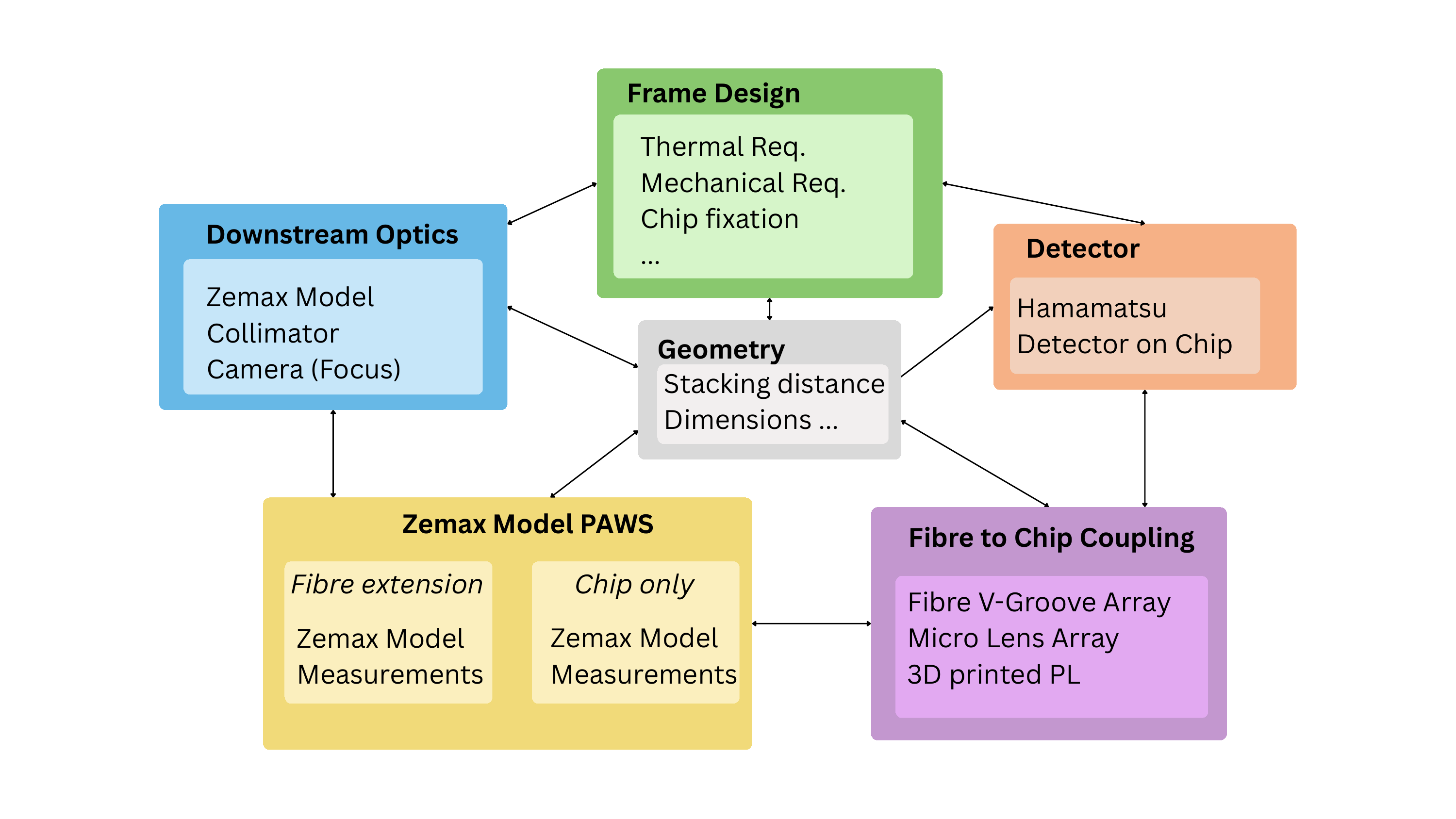}
    \caption{Linked work packages}
    \label{fig:LinkedWorkPackages}
\end{figure}

The development of \gls{cawsmos} contains multiple different work packages. At the heart of \gls{cawsmos} stands the design of the photonic chips \cite{Hernandez2023}. Equally important are the frame design to ensure alignment and spacing of the different chips, as well as the fibre to chip coupling method and the downstream optics designed in Zemax. All work packages are interlinked and cannot be treated in isolation as illustrated in Fig. \ref{fig:LinkedWorkPackages}. But the focus of this paper lies on the validation and extension of the Zemax model of \gls{paws}. The Zemax model plays an important role as it offers the opportunity to evaluate modifications in the setup without changing the physical setup and therefore allow for faster iterative learning. To actually serve as a digital twin to the experimental setup, a validation of the model was necessary which is given in Section \ref{sec:Validation}. Only then could the first extension be made as shown in Section \ref{sec:Extension} and used for the planning of experiments on the physical setup in Section \ref{sec:Results}.

\FloatBarrier

\section{Validation and extension of the Zemax model}\label{sec:ValidationAndExtension}
The validation of the Zemax model considers the position of the wavelength's signal on the detector and the size and shape of the \gls{PSF}. 

\subsection{Used data and model}
\subsubsection*{Measurement Data}\label{sec:MeasurementData}

The measured data was taken originally to calibrate \gls{paws} and allow the extraction of spectra from the detector output. The data was taken in steps of $5\text{ nm}$ from $1505 \, \text{nm}$ to $1665 \, \text{nm}$ using a tunable laser source \cite{Hernandez2023}, covering the diffraction orders 29 to 32..

\begin{figure}[h]
    \centering
    \includegraphics[width=0.8\linewidth]{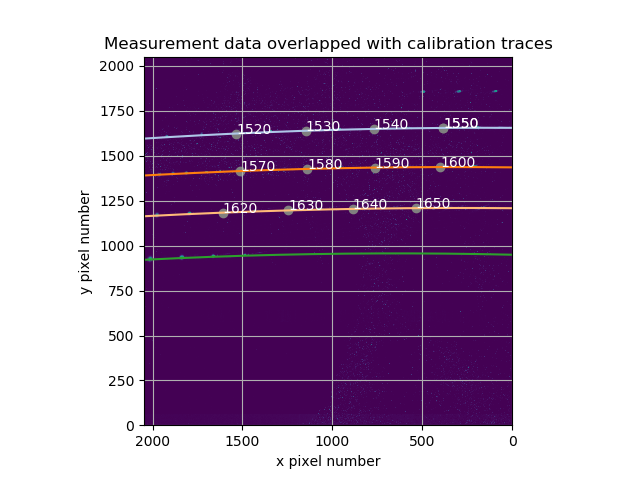}
    \caption{Measured data in $5\text{ nm}$ steps, fitted traces (solid lines) for the different orders and markers (grey) for wavelength in $ 10 \text{ nm}$ steps}
    \label{fig:CalibrationOverlap}
\end{figure}

Each spot on the detector was identified based on its intensity relative to the background and a 2D Gaussian fitted to it in order to get the  position. We divided the detector into multiple sections to treat the different orders separately, allowing parabolas to be fitted to the spots of every order as shown in Fig. \ref{fig:CalibrationOverlap}. The known corresponding wavelength for each spot makes it possible to interpolate so that the path along one of the parabolas represents a continuous spectrum in the order-specific range of wavelengths. Applied backwards, the algorithm allows to estimate the position of the \gls{PSF} of a given wavelength based on the calibration data.

\subsubsection*{Heritage Zemax Model}\label{sec:HeritageZemaxModel}

\begin{figure}[h!]
    \centering
    \includegraphics[width=0.7\linewidth]{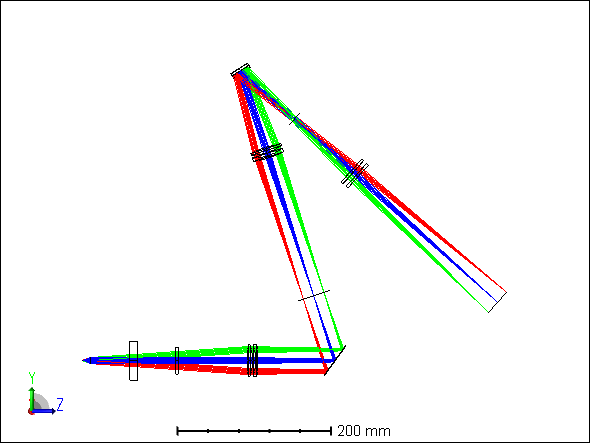}
    \caption{Original \gls{paws} Zemax model. Blue, green and red represent the fields defined at the edges and the center of the slit, respectively.}
    \label{fig:PAWSZemax}
\end{figure}

The \gls{paws} Zemax model is shown in Fig. \ref{fig:PAWSZemax}. It considers all optical elements starting with the \gls{AWG} exit slit. Hence, the path from the telescope's focal plane into and through the \gls{AWG} is not considered here. 
The \gls{AWG} chip serves as the primary dispersive element and is located in the lower left corner while the detector is positioned on the right. Within the optical path, a microscope objective with a 20x magnification, the cryostat window, the band-pass filter, multiple lenses, a mirror,  two cold-stops and a diffraction grating for cross-dispersion form the free space optics \cite{Hernandez2020}. 

The light input for the Zemax model, is a representation of the output slit of the \gls{AWG}. It is originally set up as multiple fields at different angles and positioned $10\text{ mm}$ behind the coordinate system's origin, resulting in the outer fields hitting the origin at $\pm0.7\text{ mm}$ matching the size of the output slit being $l_{slit}=1.4\text{ mm}$\cite{Hernandez2023}. Each field therefore can be interpreted as an output waveguide of an un-diced chip or simply a position on the diced \gls{AWG} slit. 

\begin{figure}
\centering
\begin{subfigure}{0.49\textwidth}
    \centering
    \includegraphics[width=0.9\linewidth]{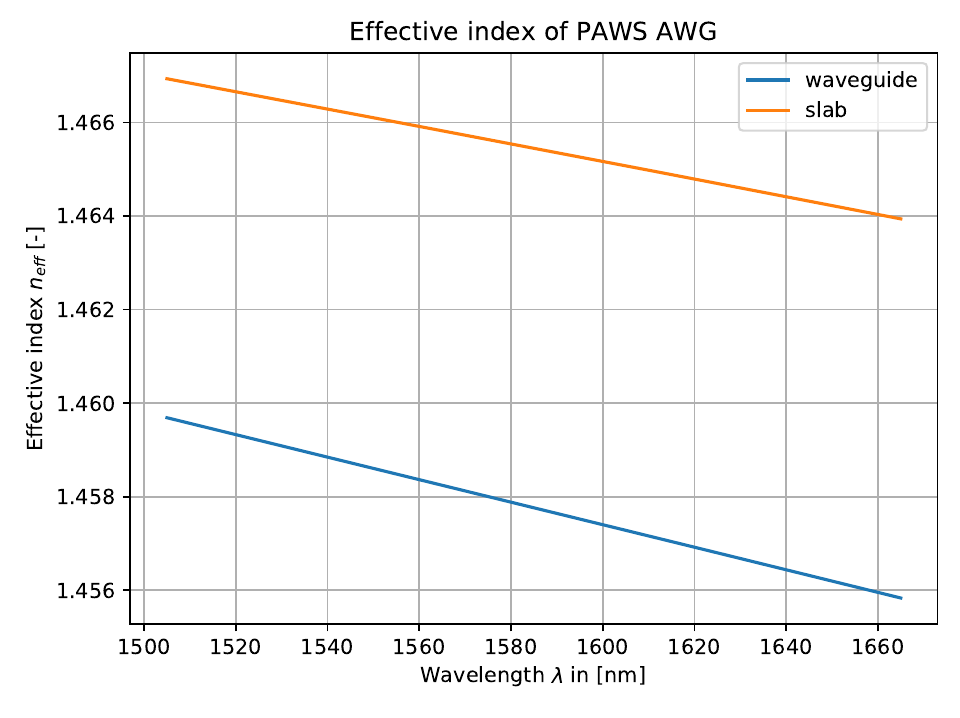}
    \caption{Effective index for waveguide (blue) and slab material (orange) for the \gls{AWG} used in \gls{paws}}
    \label{fig:neff}
\end{subfigure}%
\hfill
\begin{subfigure}{.49\textwidth}
  \centering
  \includegraphics[width=.9\linewidth]{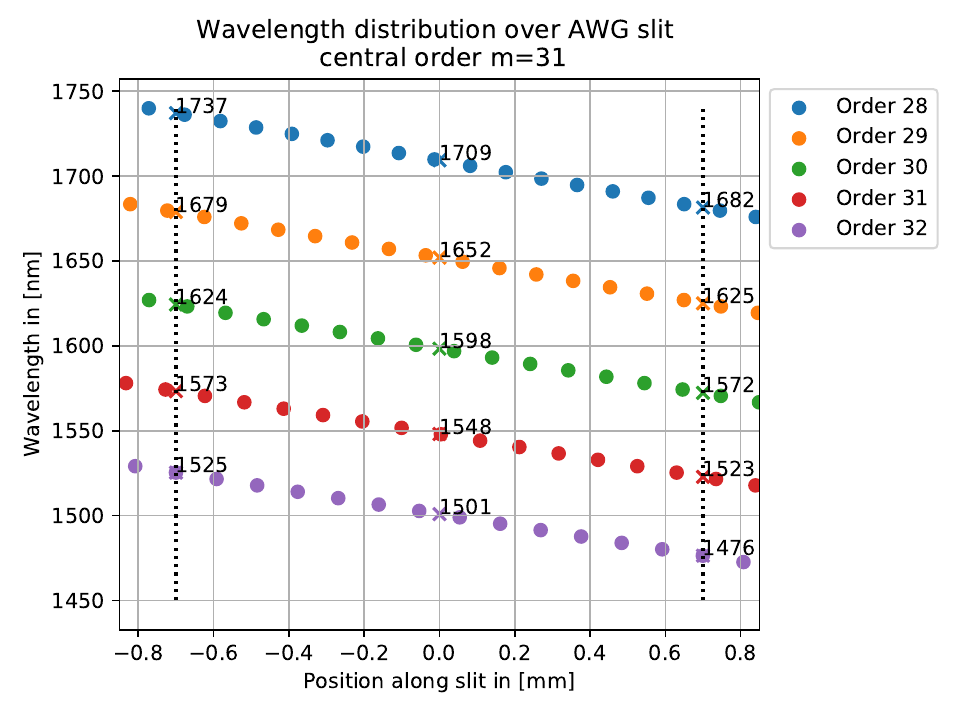}
  \caption{Positions of wavelengths at the output of the slit of the \gls{paws} AWG. The dotted lines give the slit edges.}
  \label{fig:WavelengthDistributionOverSlit}
\end{subfigure}
\caption{Determining of the positions of the individual wavelengths on the slit based on effective indices in the waveguides and slab material}
\label{fig:PositionDetermination}
\end{figure}

The investigated wavelengths are set in the wavelength-range of \gls{paws} between $1480 \text{ nm}$ and $1740\text{ nm}$\cite{Hernandez2023}. In Zemax, all fields launch all wavelengths. This is not physical. Each slit position yields only a specific wavelength per diffraction order as given by the grating equation:

\begin{align}
    m\lambda -n_{\text{eff, wg}}\cdot \Delta L= d\cdot n_{\text{eff, s}}\sin{\theta} \nonumber \\
    \theta=\arcsin{\left(\frac{m\lambda-n_{\text{eff, wg}}\cdot \Delta L}{d\cdot n_{\text{eff, s}}}\right)} \label{eq:grating2}
\end{align}
Where, $m$ is the order of diffraction, $\lambda$ the wavelength, $\Delta L$ the pathlength difference between the array waveguides (wg) and $d$ the spacing of them. For the waveguides and the slab (s) material in the \gls{fpz} the effective index is $n_{\text{eff}}$ and shown in Fig. \ref{fig:neff} as a function of wavelength.

The grating angle $\theta$ of a certain wavelength can be converted into a distinct slit position: 
\begin{align}
    \tan{\theta}=\frac{x}{f}  \label{eq:grating1}
\end{align}
where $f$ and $x$ are the focal length of the \gls{fpz} of the \gls{AWG} and the position on the slit, respectively. This then allows to evaluate the Zemax model for the correct field, corresponding to a position on the slit, for each individual wavelength. This yields the different wavelengths for each order and each position on the slit as shown in Fig. \ref{fig:WavelengthDistributionOverSlit}.

The different colors give the different orders, the central order being $m=31$ with the central wavelength of $\lambda_c=1550\text{ nm}$. As can be seen, the wavelength calculated to exit the slit at $x=0$ in the main order $m=31$ is $1548\text{ nm}$ and therefore slightly off the design central wavelength. The origin of this $2 \text{ nm}$ offset is assumed to stem from the idealized and simplified approach of calculating the expected wavelength at a certain position. However, it does not further affect the validation. 

Despite five fields being defined in the original \gls{paws} Zemax model, only the central and the outermost fields shall be investigated in the following. For these three fields, hence positions on the slit, the wavelengths for each used order were determined. It must be noted here that Zemax only takes wavelengths rounded to full nm therefore, they are not exact, but still sufficient to the purpose of linking the Zemax model with the experimental setup.

\subsection{Validation of the single chip Zemax model} \label{sec:Validation}
\subsubsection*{Position}
\begin{figure}
    \centering
    \begin{subfigure}{0.49\textwidth}
        \centering
        \includegraphics[width=1\linewidth]{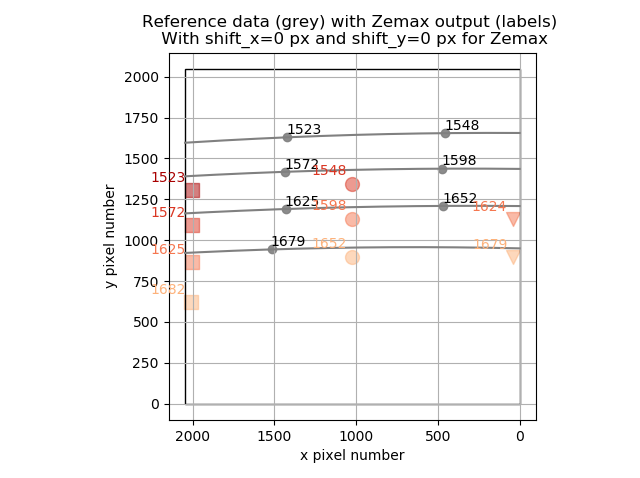}
        \caption{Unshifted}
        \label{fig:ZemaxPawsOverlapNotShifted}
    \end{subfigure}
    \hfill
    \begin{subfigure}{0.49\textwidth}
        \centering
        \includegraphics[width=1\linewidth]{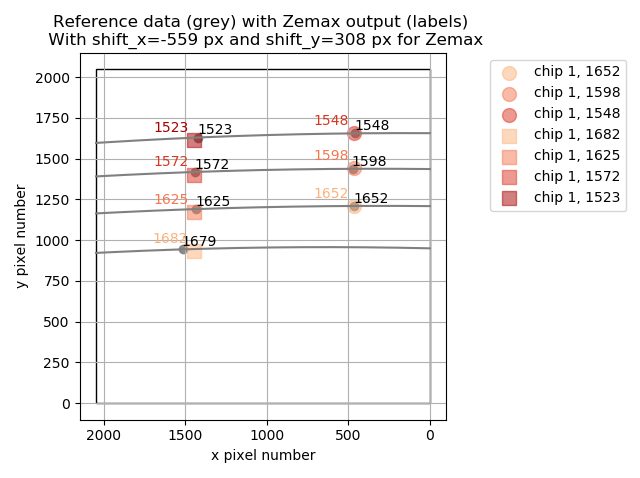}
        \caption{Shifted}
        \label{fig:ZemaxPawsOverlap}
    \end{subfigure}
    \caption{Overlap of Zemax model (reddish markers) with measured data from \gls{paws} (grey) and traces of the individual orders (solid lines)}
\end{figure}

We exported the data from Zemax via a spot diagram and used the image coordinate as the position of the different wavelengths on the detector. Reducing the \gls{PSF} to a single coordinate is a simplification applied here, as the focus lies on the position, not yet the \gls{PSF}. In the following this single coordinate \gls{PSF} will be referred to as a spot, following the convention of Zemax. Filtering the positions of the wavelengths' spots according to their assigned field yields a position diagram that gives what would be expected to be the Output of \gls{paws} if the physical setup and the Zemax model were identical. This is shown in Fig. \ref{fig:ZemaxPawsOverlapNotShifted} along with the spots as given by the measurement data and the corresponding traces of the individual orders.

It can be seen, there is a mismatch between the measured and the modelled data. The Zemax model \cite{Hernandez2020} expects $\lambda_0$ to be mapped onto the screen at $x,y=[5.64 \text{ mm}, -0.017 \text{ mm}]$, which is almost identical with the simulation result in \cite{Hernandez2023} $x,y=[5.62\text{ mm}, -0.02\text{ mm}]$. The measurements done for calibration, however place $\lambda_0$ at  $x, y=[11.54\text{ mm}, 11.34\text{ mm}$]. 

To allow better comparability and easier handling, from here on, positions are preferably given in pixel numbers. Using Eq.\ref{eq:coordinateTransform} positions can be converted from positions in mm to positions in pixel numbers and vice versa. The detector length being $l_{mm}=36.8\text{ mm}$ and $l_{\text{p}}=2048$ \cite{Hernandez2023}. This yields the positions as shown in \mbox{Table \ref{tab:positions}}. The comparison between results of the simulation from \cite{Hernandez2023} and the Zemax model show that they are almost identical with deviations $<1$ px. This bolsters the Zemax model and the data handling as described above.

\begin{align}
	{\text{val}_\text{pix}}_{\text{x}}^{\text{y}}= \pm \frac{l_{\text{pix}}}{l_\text{mm}}\cdot \text{val}_\text{mm} + \frac{l_\text{pix}}{2} \label{eq:coordinateTransform}
\end{align}

\begin{table}[h]
    \centering
    \caption{Expected and measured positions on the detector for $\lambda_0=1550\text{ nm}$}
    \begin{tabular}{lcccc}
        \textbf{Method}     & \textbf{$x_\text{mm}$} & \textbf{$y_\text{mm}$} & \textbf{$x_{\text{p}}$} & \textbf{$y_{\text{p}}$}\\
        \hline
        Simulation \cite{Hernandez2023} & 5.62 & -0.02 & 711& 1025\\
        Zemax       & 5.64 & -0.017 &711 &1025\\
        Measurement & 11.54 & 11.34 & 382 & 1655\\
        \hline
    \end{tabular}

    \label{tab:positions}
\end{table}

According to \cite{Hernandez2023} the deviation of the measurement data from simulations is mostly due to decentering. Therefore, to align the data from the Zemax model with the measured data, a simple shift is sufficient. We determined the necessary shift by minimizing the mean residual distance. The result is shown in Fig. \ref{fig:ZemaxPawsOverlap}. 

From the reddish markers representing the positions as given by Zemax, it can be seen that the light from one edge of the \gls{AWG} slit misses the detector due to the applied shift. The remaining spots coincide with their counterparts from measurement within less than 28 pixel in x and less than 18 pixel in y direction. While the residuals in x do not show any systematics, the residuals in y are close to zero for wavelengths originating from the center of the slit and close to 18 px for wavelengths originating from the edges of the slit. Considering the wavelengths missing the detector, we find that they show a similar residual but with different sign. This indicates a tilt in the cross-disperser of \gls{paws} which is not accounted for in Zemax.

\FloatBarrier
\subsubsection*{\gls{PSF}}
In addition to the position, the so far neglected size and shape of the \gls{PSF} is of interest to validate the model. 
Especially with regard to the necessary distance between chips to avoid overlap and therefore being unable to resolve and distinguish them.
To generate Zemax data at the same wavelengths as the measurement set, we created another light source in Zemax and moved it across the slit to the positions calculated for each wavelength based on Eq. \ref{eq:grating1} and \ref{eq:grating2}. We then used the Huygens \gls{PSF} function to simulate and export the \gls{PSF}.

\begin{figure}

\begin{subfigure}{0.49\textwidth}
    \centering
    \includegraphics[width=1\linewidth]{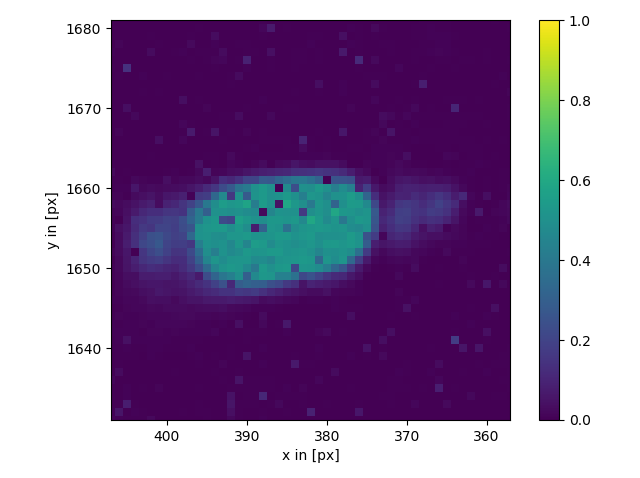}
    \caption{Measured \gls{PSF}}
    \label{fig:MeasuredPSF}
\end{subfigure}
\hfill
\begin{subfigure}{0.49\textwidth}
  \includegraphics[width=1\linewidth]{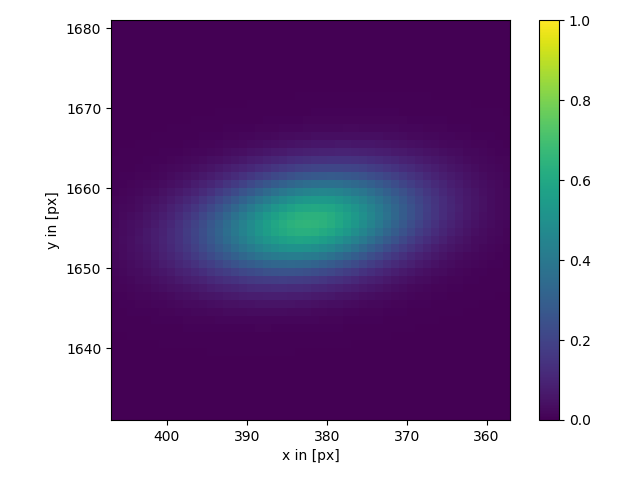}
  \caption{2D Gaussian fitted to \gls{PSF}}
  \label{fig:Measured2DGaussianPSF}
\end{subfigure}

\vspace{0.5cm}

\begin{subfigure}{0.49\textwidth}
  \includegraphics[width=1\linewidth]{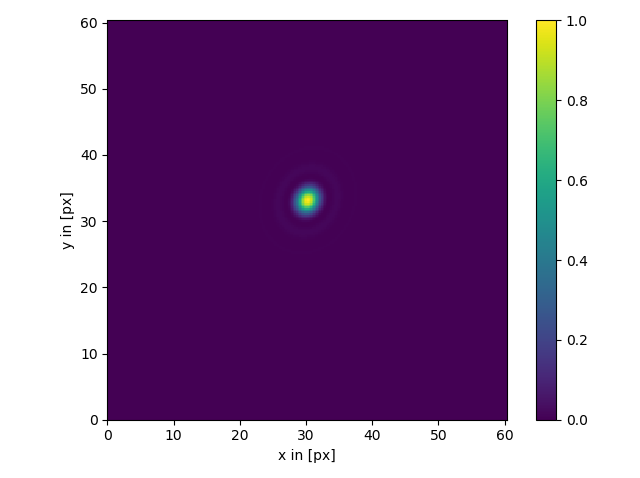}
  \caption{Zemax Huygens \gls{PSF}}
  \label{fig:Zemax2DHuygensPSF}
\end{subfigure}
\hfill
\begin{subfigure}{0.49\textwidth}
  \includegraphics[width=1\linewidth]{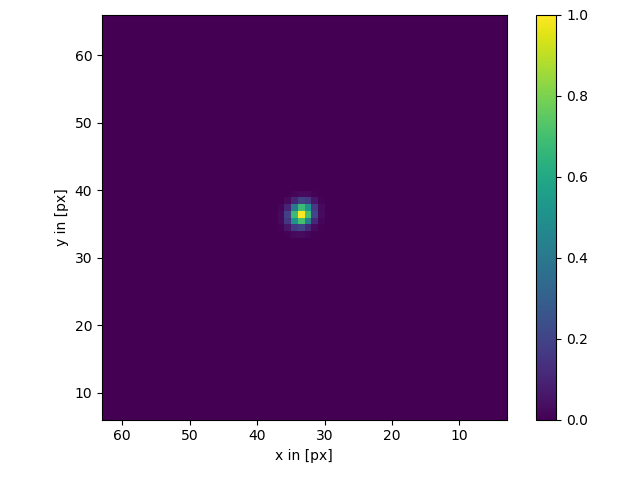}
  \caption{2D Gaussian fitted to Zemax Huygens \gls{PSF}}
  \label{fig:Zemax2DGaussianPSF}
\end{subfigure}

\caption{\gls{PSF} as measured / analyzed and fitted for measured data and Zemax simulation at $\lambda_c=1550 \text{ nm}$. Different positions come from different reference frames, but all is scaled to the pixels size of the actual detector and a window of 60x60 pixels to make the different plots comparable. The pixel size in Zemax being finer causes the original Huygens \gls{PSF} to be smoother than the fitted. The scaling remains unaffected by this.}
\label{fig:PSF_Fitting}
\end{figure}

An example for both datasets as measured / simulated and fitted is given in Fig. \ref{fig:PSF_Fitting}. 
For comparability, both \gls{PSF} were fitted with a 2D Gaussian. The \gls{fwhm} in both directions measured in px were then used to determine the geometric mean for the size and the eccentricity as indicator for distortion as given in Eq. \ref{eq:size} and \ref{eq:ecc}. To scale the \gls{PSF} from the pixel and image size in Zemax, the frame was first converted to mm and then to proper pixel number using Eq. \ref{eq:coordinateTransform}. 

\begin{align}
    \text{FWHM}=\sqrt{\text{FWHM}_x^2\cdot \text{FWHM}_y^2} \label{eq:size}\\
    e= \sqrt{1-\frac{\text{min}(\text{FWHM}_\text{x}, \text{FWHM}_\text{y})^2}{\text{max}(\text{FWHM}_\text{x}, \text{FWHM}_\text{y})^2}} \label{eq:ecc}
\end{align}

The evaluation is given in Tab. \ref{tab:PSF}. The upper and lower values giving the maximum and minimum deviation from the given mean calculated over all wavelengths.

\begin{table}[h]
    \centering
        \caption{Evaluation of \gls{PSF} in measurement and Zemax dataset via mean \gls{fwhm} in both direction as well as geometric \gls{fwhm} and the eccentricity. All are given in \gls{fwhm} in [px].}
    \begin{tabular}{lcccc}
        Data type & $\text{FWHM}_x$ & $\text{FWHM}_y$ & $\text{FWHM}$ & Eccentricity\\
        \hline
        Measurement & $11.23$ & $14.78$ & $13.17^\dscript{+5.11}{-5.52}$ & $0.7^\dscript{+0.22}{-0.28}$\\
        Zemax & $2.6$ & $2.78$& $2.68^\dscript{+0.16}{-0.16}$ & $0.48^\dscript{+0.01}{-0.01}$\\
        \hline
    \end{tabular}
    \label{tab:PSF}
\end{table}

It is obvious that the \gls{fwhm} is much larger in any direction and that the eccentricity is larger too for the measured dataset. This is because Zemax only models the optical path starting from the exit of the \gls{AWG} and under the assumption that the slit positions behave like point sources as shown in Section \ref{sec:MeasurementData}. Therefore, Zemax is missing the \gls{AWG}'s own spectral response  which is subject to aberrations in the chip such as defocal aberrations \cite{Stoll2021} and the finite number of waveguides. In addition, the interface from the chip to free-space optics and fiber to chip coupling is not modeled. However, a distinct analysis of the individual aspects' contributions is beyond the scope of this work.

\subsection{Extension of the Zemax model} \label{sec:Extension}
Taking the next step from a mere partial validation to the actual use of the Zemax model as a tool for the development of \gls{cawsmos} leads to the extension of the current model by additional sources.

\subsubsection*{Extension by fiber}
As a proof of concept we first extended the model by a single fiber of the type SMF-28. We assumed the fiber to be mounted directly on the chip with no gap in between. The distance between the point sources in Zemax results from the dimensions of the chip and the fiber which are given in Tab. \ref{tab:dimensions}. The vertical distance between the sources then becomes:

\begin{align}
    \Delta h=\frac{d_\text{chip core}}{2}+t_{\text{chip cladding}}+\frac{d_\text{coating fiber}}{2} \label{deltahfiber}
\end{align}
 where $t$ identifies a thickness and $d$ a diameter. Due to the definition in \cite{Hernandez2023} the cladding of the chip as given in Tab. \ref{tab:dimensions} already includes the thickness of the core. Therefore, Eq. \ref{deltahfiber} yields $\Delta h=136 \text{ } \mu$m.
 
\begin{table}
    \centering
        \caption{Dimensions of SMF28 fiber \cite{Thorlabs} and \gls{AWG} chip \cite{Hernandez2023} in $\mu$m}
    \begin{tabular}{lccc}
        \textbf{Component} & \textbf{Core} & \textbf{Cladding / Coating} & \textbf{Substrate + buffer}\\
        \hline 
        Fiber & 8.2 & 242 & -\\
        Chip & 3.4 & 15 & 678\\
        \hline
    \end{tabular}
    \label{tab:dimensions}
\end{table}

\begin{figure}
\centering
\begin{subfigure}{0.48\textwidth}
    \centering
    \includegraphics[width=0.9\linewidth]{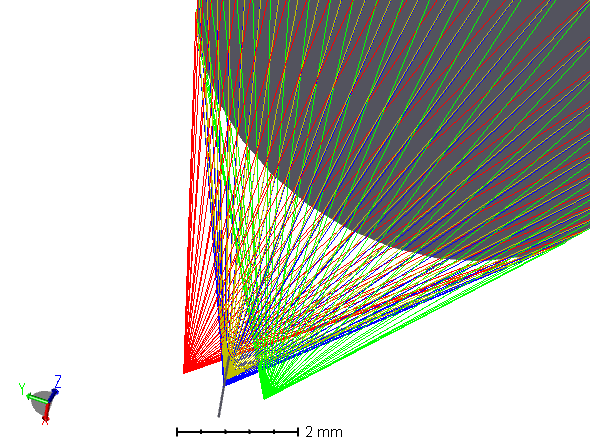}
    \caption{Chip sources in red, blue and green, single fiber extension in yellow.}
    \label{fig:ZemaxPlusFibre}
\end{subfigure}%
\hfill
\begin{subfigure}{.48\textwidth}
  \centering
  \includegraphics[width=.9\linewidth]{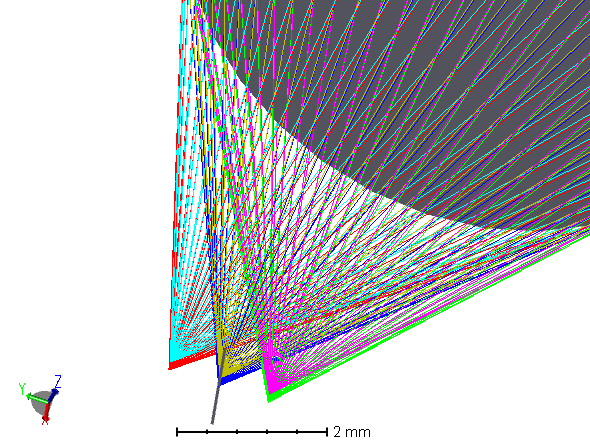}
  \caption{Original chip in red, blue and green, second chip extension in light blue, yellow and pink}
  \label{fig:ZemaxPlusChip}
\end{subfigure}
\caption{Modelling of the AWG exit slit in Zemax}
\label{fig:ZemaxAWGExit}
\end{figure}

To model $\Delta h$ orthogonal to the plane of the chip we changed the definition of the fields from Angle to Object Height since we assume the fiber being parallel to the chip's plane. Along the slit the fiber is placed in the middle as shown in Fig. \ref{fig:ZemaxPlusFibre}. To represent the first step towards a second chip the assigned wavelength is therefore $\lambda_c$. The model yields a spot in a position similar to that of the chip's central field at $\lambda_c$ but shifted in vertical direction of about $3.4$ mm.

\subsubsection*{Extension by a second chip}
\begin{figure}
    \centering
    \includegraphics[width=0.75\linewidth]{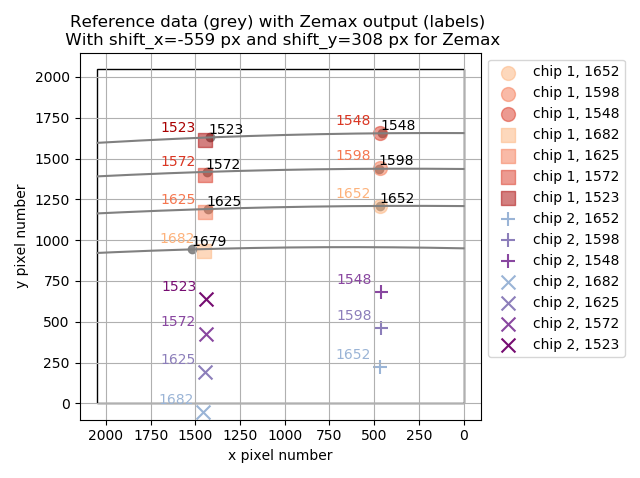}
    \caption{Zemax calculated spot positions for two stacked chips with applied shift. For the second chip (blue-purple), the spot for $1682 \text{ nm}$ hits the image plane next to the detector (solid black line)}
    \label{fig:ZemaxDoublechip}
\end{figure}

Similar to the fiber extension, a second chip can be added to the model using multiple point sources. The distance here is again given by the dimensions of the chip. Assuming the chips are identical:
\begin{align}
    \Delta h=t_\text{chip core}+t_\text{cladding}+t_\text{buffer}+t_\text{substrate}
\end{align}
so that $\Delta h=696.4 \, \mu \text{m}$ which equals the full thickness of a chip. The second chip is modeled the same way as the original chip with one field on each slit end and one in the middle using the option Object Height to define the position. The spot positions are shown in Fig. \ref{fig:ZemaxDoublechip}.

It becomes clear that a Multi-Object spectrograph in the current setup requires a much larger detector to accommodate the entire range of wavelengths in x direction to avoid a cut-off at higher wavelengths. The full spectrum can be mapped onto about 2000 pixels. In the y direction the necessary detector size is defined by the spacing of the orders within one chip, yielding about 740 necessary pixel per order. 

\FloatBarrier
\section{Results}\label{sec:Results}
Comparing the measured data with the the Zemax model, we find that the overall positions of the spots can be aligned best with a shift of -559 px in x and 308 px in y direction, minimizing the residual mean in both directions. The residual difference is less than 28 px in x direction, showing no systematic deviation. In y direction the residual lies below 18 px and indicates a tilt of the cross disperser in the experimental setup. This requires further investigation. However, relative to the \gls{fwhm} size of 20.08 px in the measured dataset, the residual is still within one resolution element.

While the position of the \gls{PSF} can be aligned well, the size and shape of the \gls{PSF} is very different in the model than in the measurement data as given in Tab. \ref{tab:PSF}. It shows that Zemax underestimates the \gls{PSF} size by a factor of 5. This is due to the limited number of aspects modeled and shows that the model is not sufficient to derive the quality of the signals of multiple chips being used in \gls{paws}.

Regardless, implementing a second chip shows that it is at least partially possible to use \gls{paws} with two chips simultaneously, considering the \gls{PSF} positions. The second chip's spectrum being shifted almost exclusively in y by 980 px. The distance between the individual orders shows that at the limits of the wavelength range, the spots of both chips risk overlap also due to the size of the \gls{fwhm} as suggested by the measurement data. Because we found Zemax' estimate of the \gls{PSF} to deviate strongly from the measurement data, no statement can be made on the quality of the second chip's mapping.

\section{Conclusions and Outlook}
In summary, the validation shows that while Zemax can be used to assess the position of a \gls{PSF} on the detector, it does not give a good indication on its qualities. 
The extensions show that a second chip stacked directly on the first one can be accommodated only partially on the detector. The next steps must therefore be to verify this result with an experiment on the actual \gls{paws} setup and the modification of the free space optics to allow for more than two chips. Furthermore the detector either needs to increase in size or the chips' output must be mapped to multiple detectors to benefit from a stacking of chips.

\acknowledgments 

The authors thank the PAWS and CAWSMOS teams at the Leibniz-Institut
f{\"u}r Astrophysik Potsdam (AIP).  Partial support for this work was provided by the PICS4SENS project, funded by the State of Brandenburg through the Investitionsbank des Landes Brandenburg (ILB), with support from the European Regional Development Fund (ERDF/EFRE), grant number 86000879.

\bibliography{lib.bib} 
\bibliographystyle{spiebib} 

\end{document}